\documentclass[11pt]{article}

\usepackage{amsmath}
\usepackage{amssymb}
\usepackage{amsthm}
\usepackage{latexsym}
\usepackage{color}
\usepackage{graphicx}
\usepackage{appendix}
\usepackage{color}
\usepackage{enumerate}
\usepackage[T1]{fontenc}
\usepackage[english]{babel}
\usepackage[table]{xcolor}

\DeclareSymbolFont{calletters}{OMS}{cmsy}{m}{n}
\DeclareSymbolFontAlphabet{\mathcal}{calletters}

\def\be{\begin{eqnarray}}
\def\ee{\end{eqnarray}}

\def\b*{\begin{eqnarray*}}
\def\e*{\end{eqnarray*}}

\newtheorem{Theorem}{Theorem}[part]

\newtheorem{Proposition}{Proposition}[part]

\newtheorem{Lemma}{Lemma}[part]
\newtheorem{Corollary}{Corollary}[part]
\newtheorem{Remark}{Remark}[part]

\makeatletter \@addtoreset{equation}{section}

\@addtoreset{Definition}{section}

\@addtoreset{Theorem}{section}

\@addtoreset{Proposition}{section}

\@addtoreset{Property}{section}

\@addtoreset{Assumption}{section}

\@addtoreset{Corollary}{section}

\@addtoreset{Lemma}{section}

\@addtoreset{Remark}{section}

\@addtoreset{Example}{section}

\@addtoreset{Condition}{section}

\def \E{\mathbb{E}}

\def \P{\mathbb{P}}
\def \Q{\mathbb{Q}}
\def \R{\mathbb{R}}

\def\={\;=\;}
\def\.{\;.}

\def\1{{\bf 1}}

\def\b*{\begin{eqnarray*}}
\def\e*{\end{eqnarray*}}

 \def\normeL2#1{\left\|{#1}\right\|_{L^2}}

 \title{VWAP execution and guaranteed VWAP\thanks{This research has been conducted with the support of the Research
Initiative ``Ex\'ecution optimale et statistiques de la liquidit\'e haute
fr\'equence'' under the aegis of the Europlace Institute of Finance. The
authors would like to thank Chris Frey (University of Alberta), Nicolas Grandchamp des Raux (HSBC France),
Jean-Michel Lasry (University Paris-Dauphine), Charles-Albert Lehalle
(CFM), Christopher Ulph (HSBC), Silviu Vlasceanu (Kepler-Cheuvreux) and Nicholas Westray (Deutsche Bank)
for the exchanges we had on the subject. Two anonymous referees also deserve to be thanked for their remarks that permitted to improve the paper in a significant manner.}}

 \author{Olivier {\sc Gu\'eant} \footnote{Universit\'e Paris-Diderot, UFR de
Math\'ematiques, Laboratoire Jacques-Louis Lions, gueant@ljll.univ-paris-diderot.fr}
 \and Guillaume {\sc Royer}\footnote{CMAP, Ecole Polytechnique Paris, guillaume.royer@polytechnique.edu.}}
             \date{}

\begin{document}

 \maketitle

 \begin{abstract}

Optimal liquidation using VWAP strategies has been considered in the literature, though never in the presence of permanent market impact and only rarely with execution costs. Moreover, only VWAP strategies have been studied and the pricing of guaranteed VWAP contracts has never been addressed. In this article, we develop a model to price guaranteed VWAP contracts in a general framework for market impact and we highlight the differences between an agency VWAP and a guaranteed VWAP contract. Numerical methods and applications are also provided.

\vspace{10mm}

\vspace{5mm}


\end{abstract}

\section{Introduction}

Traders or asset managers willing to sell blocks of shares are increasingly using execution algorithms. Amongst the strategies proposed by brokers, the most widely studied from an academic point of view is the Implementation Shortfall (IS) strategy. The classical modeling framework for optimal liquidation, developed by Almgren and Chriss in their seminal papers \cite{ac,ac2}, is indeed focused on IS orders benchmarked on the arrival price, that is the price at the beginning of the liquidation process. In the case of IS orders, the agent faces a trade-off between selling slowly to reduce execution costs and selling rapidly to limit the influence of price fluctuations. Although almost all the literature on optimal liquidation deals with IS orders, IS algorithms usually account for less volume than VWAP (Volume Weighted Average Price) algorithms  -- see for instance \cite{asx,ll}. The aim of traders when they choose VWAP orders is to focus on the reduction of execution costs: the order is split into smaller ones and the associated transactions occur on a pre-determined period to obtain a price as close as possible to the average price over this period (weighted by market volume).\footnote{On the contrary, when an investor chooses an IS algorithm, the execution process is fast at the beginning in order to obtain a price close to the one at inception. Execution costs are then usually higher for an IS than for a VWAP algorithm.} VWAP is also a neutral and rather fair benchmark to evaluate execution processes. Many agents are willing to trade as close as possible to the VWAP as they are benchmarked on the VWAP.\\

Although VWAP orders represent a large part of algorithmic trading, there are only a few papers about VWAP orders in the academic literature. The first important paper regarding liquidation with VWAP benchmark was written by Konishi \cite{k}. He developed a simple model and looked for the best static strategy, that is the best trading curve decided upon at the beginning of the liquidation process (his goal is to minimize the variance of the slippage with respect to the VWAP). He found that the optimal trading curve for VWAP liquidation has the same shape as the relative market volume curve when volatility and market volume are uncorrelated. He also quantified the deviation from the relative market volume curve in the correlated case. The model was then extended by McCulloch and Kazakov \cite{mck1} with the addition of a drift in a more constrained framework. McCulloch and Kazakov also developed a dynamic model \cite{mck2} in which they conditioned the optimal trajectory with perfect knowledge of the volume by the available information at each period of time. Bouchard and Dang proposed in \cite{b} to use a stochastic target framework to develop VWAP algorithms. Recently, Carmona and Li \cite{cl} also developed a model for VWAP trading in which the trader can explicitly choose between market orders and limit orders to buy/sell shares.\footnote{Our approach, consistently with the usual understanding of the Almgren-Chriss approach, does not prevent the use of limit orders. An execution algorithm is usually made of two layers: a strategic layer, which controls the risk with respect to a
benchmark (here the VWAP), and a tactical layer, which seeks liquidity inside order books, through all types of orders, and across other (lit or dark) liquidity pools. Our model is only concerned with the strategic layer: we want to obtain a trading curve that will then be followed by the trader using limit orders, market orders, etc.} Related to this literature on VWAP trading (see also \cite{h,kkm,w}), an academic literature appeared on market volume models. The papers by Bialkowski et al. \cite{bdf1,bdf2} or  McCulloch \cite{mc} are instances of such papers modeling market volume dynamics and the intraday seasonality of relative volume. However, all these papers ignore an important point: market impact. The only paper dealing with VWAP trading and involving a form of market impact is the interesting paper by Frei and Westray\footnote{This paper is particularly interesting since the authors characterize the unique way to model the volume process such that the relative market volume is independent from the total cumulated market volume.} \cite{fw}. In this paper, the price obtained by the trader is not the market price but a price depending linearly\footnote{Their assumption of linearity for execution costs is an acceptable one, although evidence proves that it is rather sublinear. It permits to obtain closed-form solutions.} on the desired volume (as in the early models of Almgren and Chriss). However, there is no permanent market impact in their model, while it plays an important role in our paper.\\

The aim of our model is to include permanent market impact (see \cite{gperm} for the framework we use) and any form of execution costs in a model for VWAP liquidation. Also, our goal is not to obtain a price as close as possible to the VWAP but rather to understand how to provide a guaranteed VWAP service while mitigating risk. In other words, we want to find the optimal strategy if we are given a certain quantity of shares and asked to deliver the VWAP over a predefined period. In addition to the optimal strategy underlying a guaranteed VWAP contract, we are interested in the price of such a contract. For that purpose, we use indifference pricing\footnote{For a general review on indifference pricing, the interested reader may refer to \cite{carmona}.} in a CARA framework, as in \cite{gis} where the author prices a large block of shares. This price for a guaranteed VWAP contract is the minimum premium the trader needs to pay to the broker so that the latter accepts to deliver the VWAP to the former. It depends on the size of the order, on liquidity and market conditions, and on the risk aversion of the broker.\\

In Section 2, we present the general framework of our model. We introduce the definition of the VWAP and the forms of market impact used in the model. The optimization criterion is defined and the price/premium of a guaranteed VWAP contract is defined using indifference pricing. In Section 3, we characterize the optimal liquidation strategy when the market volume curve is assumed to be known (deterministic case), along with the premium of the guaranteed VWAP contract. In Section 4, we focus on special cases and numerics. We show that, in the absence of permanent market impact, the optimal trading curve has the same shape as the market volume curve. We consider also the case of linear permanent market impact and quadratic execution costs that permits to get closed form solutions and to better understand the role played by permanent market impact. Finally, an efficient numerical method is provided to approximate the solution in the general case. In the last section, we extend our model to the case of stochastic volumes and we characterize the price of a guaranteed VWAP with a PDE.\\

\section{General framework}
\label{sect: general framework}
\subsection{Setup and notations}
\label{subsect:setup and notations}
We consider a filtered probability space $ \left( \Omega, \mathbb{F}, \left(\mathcal{F}_t \right)_{t \geq 0},\P\right)$ corresponding to the available information on the market, namely the market price and market volume of a stock up to the observation time. For $T>0$, we denote $\mathcal{P}(0,T)$ the set of $\R$-valued progressively measurable processes on [0,T].\\

We consider a trader who wants to sell $ q_0>0$ shares\footnote{A similar approach can be considered in the case of a buying order.} over the time period $[0,T]$. His liquidation strategy during this period -- hereafter denoted $v$ -- is modeled as a stochastic process belonging to the admissible set
$$ \mathcal{A}:= \left\lbrace v \in \mathcal{P}(0,T),  \int_0^T |v_t| dt \in L^\infty(\Omega), \int_0^T v_t dt= q_0, \ \P-\text{a.s.} \right\rbrace.$$
To model the trader's portfolio, we introduce a process $(q_t)_t$ that gives the number of shares remaining in the portfolio at time $t$:
$$ q_t=q_0-\int_0^t v_s ds.$$

In addition to our own volume, we introduce the instantaneous market volume process $(V_t)_t$, assumed to be progressively measurable and nonnegative. The cumulated volume from 0 up to time $t$ is denoted $ Q_t$.\footnote{In other words, $Q_t=\int_0^t V_s ds$.}

The price process of the stock is defined as a Brownian motion\footnote{The volatility parameter is not constrained to be a constant. It can be a deterministic function.} with a drift to account for market impact. Permanent market impact\footnote{Our model generalizes the linear model used in most optimal liquidation papers following Almgren and Chriss. It permits to account for the commonly observed nonlinearity of market impact. As shown in \cite{gperm}, we emphasize that there is no profitable (on average) round trip in our framework. Also, permanent market impact only depends on the initial and final positions and not on the trajectory.} is modeled by a function $ f: \R^*_+ \rightarrow \R_+$, assumed to be nonincreasing and integrable in $0$:
$$ dS_t=\sigma dW_t-f\left( \left|q_0-q_t\right| \right) v_t dt, \ \ \sigma>0. $$

The price received by the trader at time $ t$ is not $S_t$ because of temporary market impact. This temporary market impact, also referred to as execution costs, is modeled through the introduction of a function $L\in C(\R,\R_+)$ verifying:
\begin{itemize}
\item $ L(0)=0 $,
\item $ L$ is an even function,
\item $L$ is increasing on $\R_+$,
\item $L$ is strictly convex,
\item $L$ is asymptotically superlinear, that is:
$$\lim_{\rho \to + \infty }\frac{L(\rho)}{\rho} = + \infty.$$
\end{itemize}

\begin{Remark}
In applications, $L$ is often a power function, i.e. $ L(\rho)=\eta \left| \rho\right|^{1+\phi}$ with $ \phi>0$, or a function of the form $ L(\rho)=\eta \left| \rho\right|^{1+\phi} + \psi |\rho|$ with $ \phi,\psi>0$.\\
\end{Remark}

For any $v \in \mathcal{A}$, we define the cash process $X^v$ (hereafter denoted $X$ to simplify notations) by:
$$ X_t = X_t^v =\int_0^t \left( v_s S_s -V_s L\left( \frac{v_s}{V_s} \right)  \right) ds.  $$

We are interested in the following optimization problem:
\begin{align}\label{def maximisation}
\sup_{v\in \mathcal{A}} \E \left[ -\exp\left(-\gamma (X_T-q_0 \text{VWAP}_T) \right)\right],
\end{align}

where\footnote{The limiting case $\gamma=0$ corresponds to risk neutrality. Here, we choose to consider an expected utility framework with a CARA utility function. This framework boils down to a mean variance setting in the case of gaussian risks. Considering a utility function is more rigorous from an economic viewpoint, especially when it comes to pricing.} $\gamma>0$ is the absolute risk aversion parameter, and where $ \text{VWAP}_T$ stands for the volume weighted average price (VWAP) on the period $ [0,T]$, i.e.:
\begin{align}\label{def VWAP}
\text{VWAP}_T:= \frac{\int_0^T S_t V_t dt}{\int_0^T V_t dt} = \frac{\int_0^T S_t dQ_t}{Q_T}.
\end{align}

\begin{Remark}
The above definition of the VWAP can also be formulated as:
$$ \text{VWAP}_T=\int_0^T S_t dx_t,$$
where $ x_t:= \frac{Q_t}{Q_T}$. This formulation is often used in the literature but it has an important drawback as the above integral is not $ \mathbb{F}$-adapted. Its natural filtration is indeed $(\mathcal{G}_t)_t$ where:
$$ \mathcal{G}_t:= \mathcal{F}_t \vee \sigma(Q_T). $$
\end{Remark}

\begin{Remark}
In the above definition of $VWAP_T$, we did not include our own volume. An alternative definition, including our trades, is
$$ \text{VWAP}'_T:= \frac{\int_0^T S_t (V_t + v_t) dt}{Q_T + q_0}.$$
We shall see in the appendix of this article that the results we obtain with our simpler definition can be easily modified to be true in the case of the alternative definition.
\end{Remark}

The rationale for our optimization criterion \eqref{def maximisation} is the following. We consider a stock trader or an asset manager who wants to sell $q_0$ shares at the VWAP over the period $[0,T]$. For that purpose, he signs with an intermediary (typically a broker) a guaranteed VWAP contract.\footnote{This is not an agency VWAP since, here, the VWAP is guaranteed (although its value is not known \emph{ex ante}).} He gives the intermediary its $q_0$ shares at time 0 and he receives at time $T$, $q_0$ times the VWAP computed over the period $[0,T]$, minus a premium -- hereafter denoted $\pi(q_0)$ -- to compensate the intermediary for the service and the associated costs, this premium being agreed upon at time $0$. The problem we address is the problem of the intermediary: he receives $q_0$ shares and sells them over the period $[0,T]$ to obtain $X_T$ on its cash account at time $T$. Then, he gives his client $q_0 \text{VWAP}_T - \pi(q_0)$ in cash. Therefore, if the intermediary has a constant absolute risk aversion $\gamma$, his liquidation strategy is obtained by maximizing
$$\E \left[ -\exp\left(-\gamma (X_T-q_0 \text{VWAP}_T + \pi(q_0)) \right)\right] =\exp(-\gamma \pi(q_0)) \E \left[-\exp\left(-\gamma (X_T-q_0 \text{VWAP}_T) \right)\right].$$

Since $\pi(q_0)$ is agreed upon at time $0$, the objective function is simply:
$$\E \left[-\exp\left(-\gamma (X_T-q_0 \text{VWAP}_T) \right)\right].$$

Now, to decide upon the value $\pi(q_0)$, we shall compute the reservation price of the intermediary, that is the price at which the intermediary is indifferent between accepting the contract and refusing it. This approach for pricing, also called indifference pricing, leads to the following definition for the premium of a guaranteed VWAP:\footnote{The premium in percentage is then $ \frac{\pi (q_0)}{q_0 S_0} $. However, this is expressed in percentage of the initial price. In practice, some guaranteed VWAP contracts are also priced in basis points of the realized VWAP. We refer to Appendix B for this different definition of the premium and its consequences on our approach. Numerically, the two approaches provide similar figures in basis point of their respective reference price.}

$$\pi(q_0) = \frac 1\gamma \log\left(- \sup_{v\in \mathcal{A}} \E \left[ -\exp\left(-\gamma (X_T-q_0 \text{VWAP}_T) \right)\right]\right).$$

Our goal in this paper is twofold: (i) we want to solve the optimization problem, that is to find an optimal liquidation strategy $v$, and (ii) we want to find the value of $\pi(q_0)$.\\

\subsection{First properties}
\label{subsect: some easy calculations}
We derive here the key formulas for $ X_T$ and $ \text{VWAP}_T$:
\begin{Lemma}\label{Lem expression of XT}
For any $ v \in \mathcal{A}$, we have:
$$ X_T=q_0 S_0-\int_0^{q_0}F(z)dz- \int_0^T V_t L\left( \frac{v_t}{V_t} \right)  dt + \int_0^T q_t \sigma dW_t , $$
where $ F(q)=\int_0^qf(\left|z\right|)dz$
\end{Lemma}

\textbf{Proof:}\\

Integrating by parts, we have:
\begin{align*}
X_T &= \int_0^T v_t S_t dt - \int_0^T V_t L\left( \frac{v_t}{V_t} \right)dt \\
 &= \left[ -q_t S_t\right]_0^T +\int_0^T q_t dS_t -  \int_0^T V_t L\left( \frac{v_t}{V_t} \right)dt \\
 &=q_0 S_0-\int_0^T q_t f\left(\left|q_0-q_t \right| \right) v_tdt+\int_0^T q_t \sigma dW_t -  \int_0^T V_t L\left( \frac{v_t}{V_t} \right)dt .
\end{align*}
We then observe that:
$$
\int_0^T q_t f\left(\left|q_0-q_t \right| \right) v_tdt = \left[ q_t F\left( q_0-q_t\right)\right]_0^T+\int_0^T F(q_0-q_t) v_t dt =\int_0^T F(q_0-q_t) v_t dt
$$
Defining in the last integral the change of variables $ z=q_0-q_t$, we have:
$$ \int_0^T q_t f\left( \left|q_0-q_t \right| \right) v_tdt =\int_{0}^{q_0}F(z)dz. $$
Finally, we obtain:
$$ X_T=q_0 S_0-\int_0^{q_0}F(z)dz- \int_0^T V_t L\left( \frac{v_t}{V_t} \right)  dt + \int_0^T q_t \sigma dW_t.$$
\begin{flushright}$\Box$\end{flushright}

The same calculation can be made for VWAP:
\begin{Lemma}
We have:
$$ \text{VWAP}_T= S_0+\int_0^T \left( 1-\frac{Q_t}{Q_T} \right)\sigma dW_t- \int_0^T \frac{V_t}{Q_T} F(q_0-q_t)dt.$$
\end{Lemma}

\textbf{Proof:}\\

This is an integration by parts:
\begin{align*}
\text{VWAP}_T &= \frac{1}{Q_T} \int_0^T S_t V_t dt \\
&=   \left[S_t \left(\frac{Q_t}{Q_T}-1 \right) \right]_0^T- \int_0^T\left( \frac{Q_t}{Q_T}-1\right)\sigma dW_t + \int_0^T \left(\frac{Q_t}{Q_T}-1 \right)f(\left| q_0-q_t\right| )v_t dt\\
&= S_0+\int_0^T \left( 1-\frac{Q_t}{Q_T} \right)\sigma dW_t +  \left[\left(\frac{Q_t}{Q_T}-1\right) F(q_0-q_t) \right]_0^T- \int_0^T \frac{V_t}{Q_T} F(q_0-q_t)dt\\
&=S_0+\int_0^T \left( 1-\frac{Q_t}{Q_T} \right)\sigma dW_t- \int_0^T \frac{V_t}{Q_T} F(q_0-q_t)dt,
\end{align*}
where we recall that $ Q_0=0$.
\begin{flushright}$\Box$\end{flushright}

The slippage $X_T-q_0 \text{VWAP}_T$ is then given by:

\begin{Corollary}\label{slippage value}
For $ v \in \mathcal{A}$, we have:
\begin{align*}
X_T-q_0 \text{VWAP}_T=&-\int_0^{q_0}F(z)dz-\int_0^T V_t L\left(\frac{v_t}{V_t} \right)dt+q_0\int_0^T \frac{V_t}{Q_T} F(q_0-q_t) dt\\
&+ \sigma q_0 \int_0^T \left(\frac{q_t}{q_0}-\left( 1-\frac{Q_t}{Q_T}\right) \right)dW_t.
\end{align*}
\end{Corollary}

\section{The deterministic case}
\label{sect: deterministic case}
In this section we assume that the volume curve is deterministic. We also assume that $ V$ is bounded from above and from below:
$$ \underline{V} \le V_{\cdot} \le \overline{V}, \ \text{for some} \ \overline{V},\underline{V}>0 .$$

We first study the case where we restrict the set of admissible strategies to the deterministic ones:
$$ \mathcal{A}_{det}=\lbrace v \in \mathcal{A}: \ v \ \text{is} \ \mathcal{F}_0-\text{measurable} \rbrace.$$
The problem \eqref{def maximisation} is then replaced by:
\begin{align}\label{def maximisation det}
\sup_{v\in \mathcal{A}_{det}} \E \left[ -\exp\left(-\gamma (X_T-q_0 \text{VWAP}_T) \right)\right],
\end{align}
In that framework, we obtain the following lemma:
\begin{Lemma}\label{Lemme 1 section 3}
For any $ v\in \mathcal{A}_{det}$,  $ X_T-q_0 \text{VWAP}_T$ is normally distributed with mean
$$ -\int_0^{q_0}F(z)dz-\int_0^T V_t L\left(\frac{v_t}{V_t} \right)dt+q_0\int_0^T \frac{V_t}{Q_T} F(q_0-q_t) dt $$
and variance
$$  \sigma^2 q_0^2 \int_0^T \left(\frac{q_t}{q_0}-\left( 1-\frac{Q_t}{Q_T}\right) \right)^2dt .$$
\end{Lemma}

\textbf{Proof:}\\

The proof is straightforward, considering the formula obtained in Corollary \ref{slippage value}.\qed\\

Using the expression of the Laplace transform of a normally distributed variable, we obtain:

\begin{align*}
\sup_{v\in \mathcal{A}_{det}} &\E \left[  -\exp\left(-\gamma (X_T-q_0 \text{VWAP}_T) \right)\right]\\
= \sup_{v\in \mathcal{A}_{det}} &-\exp\left(-\gamma \left(-\int_0^{q_0}F(z)dz-\int_0^T V_t L\left(\frac{v_t}{V_t} \right)dt+q_0\int_0^T \frac{V_t}{Q_T} F(q_0-q_t) dt\right. \right.\\
& \left. \left. - \frac{\gamma}{2}\sigma^2 q_0^2 \int_0^T \left(\frac{q_t}{q_0}-\left( 1-\frac{Q_t}{Q_T}\right) \right)^2dt \right) \right).
\end{align*}
Our optimization problem consequently boils down to:
\begin{align}\label{minimization pb det}
\inf_{v \in \mathcal{A}_{det}} \int_0^T V_t L\left(\frac{v_t}{V_t} \right)dt-q_0\int_0^T \frac{V_t}{Q_T} F(q_0-q_t) dt+ \frac{\gamma}{2}\sigma^2 q_0^2 \int_0^T \left(\frac{q_t}{q_0}-\left( 1-\frac{Q_t}{Q_T}\right) \right)^2dt.
\end{align}

In order to obtain existence and uniqueness, we use the classical framework developed in \cite{cs}. We denote $ AC_{q_0,0}(0,T)$ the set of absolutely continuous arcs $ q: [0,T] \rightarrow \R $ such that $ q(0)=q_0$ and $ q(T)=0$ and we define the application $ \mathcal{I} : AC_{q_0,0}(0,T) \to \mathbb{R} $ by:
$$\mathcal{I}(q) = \displaystyle\int_0^T \left( V_t L\left(\frac{\dot{q}(t)}{V_t} \right)-q_0\frac{V_t}{Q_T} F(q_0-q(t)) +\frac{\gamma}{2}\sigma^2 q_0^2\left(\frac{q(t)}{q_0}-\left( 1-\frac{Q_t}{Q_T}\right) \right)^2  \right) dt.$$
We shall denote $ \ell(t,q,v):= V_t L\left(\frac{v}{V_t} \right)-q_0\frac{V_t}{Q_T} F(q_0-q) +\frac{\gamma}{2}\sigma^2 q_0^2\left(\frac{q}{q_0}-\left( 1-\frac{Q_t}{Q_T}\right) \right)^2 $ for short.\\

In order to prove the existence of an optimal strategy, we first show the technical lemma:
\begin{Lemma}\label{lemme technique existence}
$ \ell$ verifies the three following assertions:
\begin{itemize}
\item[$\rm{(i)}$] $ \ell$ is convex with respect to the third variable;
\item[$\rm{(ii)}$]  there exists $ c_0 \geq 0$ and $ \theta : \R_+ \rightarrow \R_+$ such that $ \theta(v)/v \rightarrow +\infty$ as $ v\rightarrow +\infty$ and
$$ \forall t \in [0,T], q\in \R, \ell(t,q,v) \geq \theta(\left| v\right| )-c_0;$$
\item [$\rm{(iii)}$] for all $ r>0$ there exists $ C(r) >0 $ such that
$$ \left| \ell(t,q,v)-\ell(t,\tilde{q},v)\right| <C(r) \omega( \left| q-\tilde{q}\right| )\theta(\left| v\right| )$$
for all $ t\in[0,T]$, $ q$, $ \tilde{q}\in [-r,r] $, $ v\in \R$, where $ \omega : \R_+ \rightarrow \R_+$ is a modulus of continuity.
\end{itemize}
\end{Lemma}
\textbf{Proof:}\\

(i) follows from the convexity of $ L$ and the positivity of $ V$.\\

For (ii), we first see that for any $ t\in[0,T]$ and $ q\in \R$, we have:
$$ -q_0\frac{V_t}{Q_T} F(q_0-q) +\frac{\gamma}{2}\sigma^2 q_0^2\left(\frac{q}{q_0}-\left( 1-\frac{Q_t}{Q_T}\right) \right)^2 \geq g(q) $$
where
$$ g(q):=-q_0\frac{\bar{V}}{Q_T} \left| F(q_0-q)\right| +\frac{\gamma}{2}\sigma^2q_0^2 \inf_{a\in [0,1]} \left(\frac{q}{q_0} -a \right)^2.$$
$ g$ is continuous with $ g(q)\underset{q\rightarrow +\infty}\longrightarrow +\infty $ and $ g(q)\underset{q\rightarrow -\infty}\longrightarrow +\infty $, so that there exists $ c_0$ such that $ g \geq c_0$ on $ \R$.\\
We next define $ \theta(v):=\underline{V}L\left( \frac{v}{\overline{V}}\right)$. By the superlinearity of $ L$, we have $ \theta(v)/v \rightarrow +\infty$ as $ v \rightarrow+\infty$. Finally, since $ L$ is even, we have for all $(t,q,v) \in [0,T] \times \R \times \R$:
$$ \ell(t,q,v) \geq \theta(\left| v\right| )-c_0 .$$
For (iii), we have for all $(t,q,\tilde{q},v) \in [0,T] \times \R\times \R \times \R$:
\begin{align}\label{inegalite l}
\left|\ell(t,q,v)-\ell(t,\tilde{q},v) \right| \le C \left( \left| F(q_0-q)-F(q_0-\tilde{q})\right| + \left| \frac{q^2}{q_0^2}-\frac{\tilde{q}^2}{q_0^2} \right| +\frac{2}{q_0} \left| q-\tilde{q}\right|  \right)
\end{align}
where $ C$ is a constant uniform in $ t$ and $ v$. Since $ f$ is nonincreasing on $\R_+$, we have for any $ a \le b \in R$: $ 0 \le \int_a^bf(\left| z\right| )dz \le \int_{(a-b)/2}^{(b-a)/2}f(\left| z\right| )dz$ and then:
\begin{align}\label{inegalite F}
 \left| F(q_0-q)-F(q_0-\tilde{q})\right|  \le \int_{-\left| q-\tilde{q}\right| /2}^{\left| q-\tilde{q}\right| /2} f(\left| z\right| )dz=:\omega_1(\left| q-\tilde{q}\right| ).
\end{align}
Combining \eqref{inegalite l} and \eqref{inegalite F}, we obtain:
$$ \left| \ell(t,q,v)-\ell(t,\tilde{q},v)\right|  \le C(r) \omega(\left| q-\tilde{q}\right| )$$
where $ C(r):=C(1+\frac{2r}{q_0^2}+\frac{2}{q_0})$ and $ \omega(r):= \omega_1(r)+r$.
\begin{flushright}$\Box$\end{flushright}

\begin{Theorem}\label{existence}
There exists a unique minimizer $ q^*$ of $ \mathcal{I}$ in $ AC_{q_0,0}(0,T)$. Moreover we have that $ q^*(t) \le q_0$ for all $ t\in [0,T]$.
\end{Theorem}
\textbf{Proof:}\\

We divide the proof in three steps.\\

\textit{Step 1:} We first show that any strategy can be improved by considering a new one taking values in $ (-\infty,q_0]$. Indeed for $ q \in AC_{q_0,0}(0,T) $, we define $ \tilde{q}$ by $ \tilde{q}(t):=\min(q_0,q(t))$. We have $ \tilde{q} \in AC_{q_0,0}(0,T)$ and $ \dot{\tilde{q}}(t)=\dot{q}(t){\bf{1}}_{q(t)<q_0}$. Therefore, since $L$ is even and increasing on $\R_+$, we have:
 $$ L\left( \frac{\dot{\tilde{q}}(t)}{V_t}\right) = L\left( \frac{|\dot{\tilde{q}}(t)|}{V_t}\right) \le L\left( \frac{|\dot{q}(t)|}{V_t}\right) = L\left( \frac{\dot{q}(t)}{V_t}\right)$$
Also, since $ F$ is odd and nondecreasing, we have for any $ t\in [0,T]$:
 $$- F(q_0-\tilde{q}(t)) \le - F(q_0-q(t)).  $$
 Eventually, we have that
 $$ \left(\frac{\tilde{q}(t)}{q_0}- \left(\frac{Q_T-Q_t}{Q_T} \right) \right)^2 \le \left(\frac{q(t)}{q_0}- \left(\frac{Q_T-Q_t}{Q_T} \right) \right)^2, $$
and then:
$$ \mathcal{I}(\tilde{q} ) \le \mathcal{I}( q), $$
with strict inequality whenever $ \tilde{q} \neq q$.\\

\textit{Step 2:} We then show the uniqueness of the minimizer. Let us consider $ q_1$ and $ q_2$ two minimizers of $ \mathcal{I}$ such that $ q_1 \neq q_2$. We know, using Step 1, that $ q_1$ and $ q_2$ take values in $ (-\infty,q_0]$. We next define $ q$ by $ q(t):=\frac{q_1(t)+q_2(t)}{2}$. By convexity of $ L$, convexity of $ q\rightarrow -F(q_0-q)$ on $ (-\infty,q_0]$ and strict convexity of $ q \rightarrow \left(q-a \right)^2$, we have $ \mathcal{I}(q) <\frac{1}{2}\mathcal{I}(q_1)+\frac{1}{2} \mathcal{I}(q_2)$, which contradicts the optimality of $ q_1$ and $ q_2$.\\

\textit{Step 3:} The existence of the solution follows from Theorem 6.1.2 in \cite{cs}, where we show in Lemma \ref{lemme technique existence} that $ \ell $ verifies the three required conditions.

\begin{flushright}$\Box$\end{flushright}

We shall characterize $q^*$ as the solution of the Hamiltonian system associated to the optimization problem.\\
The function $\ell$ is not convex and so the classical results of \cite{r1} and \cite{r2} cannot be applied. Since the optimal solution $ q^*$ must verify $ q^* \le q_0$, we modify the problem and introduce $ \tilde{F}(q):=F(q) {\bf{1}}_{q \geq 0}-\infty {\bf{1}}_{q <0}$. We then define the associated function $ \tilde{\ell}$ on $ [0,T] \times \R \times \R $ as an extended real-valued function by:

$$ \tilde{\ell}(t,q,v):=V_t L\left(\frac{v}{V_t} \right)-q_0\frac{V_t}{Q_T} \tilde{F}(q_0-q) +\frac{\gamma}{2}\sigma^2 q_0^2\left(\frac{q}{q_0}-\left( 1-\frac{Q_t}{Q_T}\right) \right)^2. $$
Then, using Theorem \ref{existence}, we clearly see that the problem \eqref{minimization pb det} is identical to the problem:
\begin{align}\label{minimization pb det convex}
\inf_{v \in \mathcal{A}_{det}} \int_0^T V_t L\left(\frac{v_t}{V_t} \right)dt-q_0\int_0^T \frac{V_t}{Q_T} \tilde{F}(q_0-q_t) dt+ \frac{\gamma}{2}\sigma^2 q_0^2 \int_0^T \left(\frac{q_t}{q_0}-\left( 1-\frac{Q_t}{Q_T}\right) \right)^2dt,
\end{align}
and an optimal strategy for one of the problem, is also an optimal strategy for the other problem.\\

To solve this problem, we use the technics developed in \cite{r1} and \cite{r2}. $ \tilde{\ell}(t,\cdot,\cdot)$ is indeed a convex function, taking values in $ \R \cup \lbrace+\infty\rbrace$. We note that conditions (B), (C) and (D) enumerated in \cite{r1} are also satisfied, given the assumptions on $ V_\cdot$ and the assumptions on $L$.\\

We now introduce the Legendre transform of $ L$ defined by $ H(p)=\sup_{\rho \in \R} \rho p - L( \rho)$. Since $ L$ is strictly convex, we recall that $ H$ is a $ C^1$ function.\\

Using this Legendre transform, we have the following characterization of $q^*$:

\begin{Proposition}[Hamiltonian system]\label{prop:deterministic hamiltonian}
For $ q\in AC_{q_0,0}(0,T)$, we have equivalence between the two following assertions:
\begin{itemize}
\item [$\rm{(i)}$] $ q=q^*$;
\item [$\rm{(ii)}$] there exists $ p \in AC(0,T)$ such that for all $ t\in[0,T]$:
$$\left\lbrace
\begin{array}{lcl}
\dot{p}(t) &=& \gamma \sigma^2 \left( q(t)-q_0\left( 1 - \frac{Q_t}{Q_T}\right)\right)+q_0\frac{V_t}{Q_T}f(\left|q_0-q(t)\right| )\\
\dot{q}(t) &=& V_t H'(p(t))
\end{array}\right. \ \ \ q(0)=q_0, \ \ q(T)=0.$$
\end{itemize}

\end{Proposition}
\textbf{Proof}:\\

The Hamiltonian of the system is:
$$ \mathcal{H}(t,q,p):=V_t H(p)+q_0\frac{V_t}{Q_T} \tilde{F}(q_0-q)-\frac{\gamma}{2}\sigma^2 q_0^2\left(\frac{q}{q_0}-\left( 1-\frac{Q_t}{Q_T}\right) \right)^2.$$

The characterization of $q^*$ given in the Theorem 6 of \cite{r1} and its corollary is:
\begin{align*}
\dot{q}(t) &\in \partial^-_p \mathcal{H}(t,q(t),p(t))\\
\dot{p}(t) &\in \partial^-_q \left(-\mathcal{H}\right)(t,q(t),p(t)),
\end{align*}
where $\partial^-$ stands for the subdifferential.\\

Given the expression for $\mathcal{H}$, only $\partial^-_q \left(-\mathcal{H}\right)(t,q(t),p(t))$ may not be a singleton made of a real number, when $q(t) = q_0$. If $\lim_{x \to 0^+} f(x)$ is finite, then the expression given in the Proposition is obtained by straightforward computation. If  $\lim_{x \to 0^+} f(x) = +\infty$, then $\partial^-_q \left(-\mathcal{H}\right)(t,q_0,p(t)) = \lbrace + \infty \rbrace$ and the expression in the Proposition is correct, giving $\dot{p}(t) = +\infty$ whenever $q(t) = q_0$.
\begin{flushright}$\Box$\end{flushright}

This hamiltonian characterization allows to get a regularity result for the optimal strategy

\begin{Corollary}
If $V_{\cdot}$ is continuous then $q \in C^1([0,T])$
\end{Corollary}

\textbf{Proof:}\\

$L$ being strictly convex, $H$ is $C^1$ and the result is a consequence of the equation $\dot{q}(t) = V_t H'(p(t))$.\qed\\

\begin{Remark}
Even when $V_\cdot$ is smooth or even constant, $q^*$ may not be $C^2$. This remark is important since most studies are using a Euler-Lagrange equation in the classical sense with the underlying assumption that the strategy is $C^2$.
\end{Remark}

We end this section by the optimality of deterministic strategies. This result was first proved in \cite{s} in the case of Implementation Shortfall (IS) liquidation strategies. We use the same method here in the case of VWAP liquidation strategies.
\begin{Theorem}\label{th deterministic strategies are optimal}
Assume that $ V_\cdot$ is deterministic, then:
$$ \sup_{v \in \mathcal{A}_{det}} \E \left[ -\exp\left(-\gamma (X_T-q_0 \text{VWAP}_T) \right)\right] =\sup_{v \in \mathcal{A}} \E \left[ -\exp\left(-\gamma (X_T-q_0 \text{VWAP}_T) \right)\right].$$
\end{Theorem}
\textbf{Proof:}\\

Let us consider $v \in \mathcal{A}$. We have:
\begin{align*}
&\E \left[ -\exp \left(-\gamma \left(X_T-q_0 \text{VWAP}_T\right) \right)\right]\\
=&-\exp\left(\gamma \int_0^{q_0}F(z)dz \right)\\
&\times \E \left[ \exp \left(\gamma \int_0^T\left( V_t L \left( \frac{v_t}{V_t}\right)-q_0 \frac{V_t}{Q_T} F(q_0-q_t) \right)dt \right)\right. \\
&\times \left.\exp\left(  -\gamma \sigma q_0 \int_0^T \left( \frac{q_t}{q_0}-\frac{Q_T-Q_t}{Q_T} \right)dW_t \right)  \right].\\
\end{align*}
We then define a new probability measure $ \Q $ by:
$$ \frac{d \Q}{d\P}:=\exp\left(  -\gamma \sigma q_0 \int_0^T \left( \frac{q_t}{q_0}-\frac{Q_T-Q_t}{Q_T} \right)dW_t -\frac{1}{2}\gamma^2 q_0^2 \sigma^2 \int_0^T \left( \frac{q_t}{q_0}-\frac{Q_T-Q_t}{Q_T} \right)^2 dt\right).$$
Since $v \in \mathcal{A} \implies q \in L^\infty(\Omega \times (0,T))$, we observe that $ \frac{d\Q }{d\P}$ indeed defines a change of probability.\\

We then have:
$$ \E \left[ -\exp \left(-\gamma \left(X_T-q_0 \text{VWAP}_T\right)  \right)\right] =-\exp\left(\gamma \int_0^{q_0}F(z)dz \right)  \E^{\Q} \left[ \exp \left( \gamma \mathcal{I}(q) \right) \right].$$

Now, $\P$-a.s., $q(\omega) \in AC_{q_0,0}(0,T)$ so that $\P$-a.s., $\mathcal{I}(q(\omega)) \ge \mathcal{I}(q^*)$. This gives:

\begin{align*}
\E \left[ -\exp \left(-\gamma \left(X_T-q_0 \text{VWAP}_T\right)  \right)\right]&= -\exp\left(\gamma \int_0^{q_0}F(z)dz \right)  \E^{\Q} \left[ \exp \left( \gamma \mathcal{I}(q) \right) \right]\\
& \le -\exp\left(\gamma \int_0^{q_0}F(z)dz \right)  \exp \left(\gamma \mathcal{I}(q^*) \right)\\
& = \sup_{v \in \mathcal{A}_{det}} \E \left[ -\exp\left(-\gamma (X_T-q_0 \text{VWAP}_T) \right)\right].
\end{align*}

Hence:
$$\sup_{v \in \mathcal{A}} \E \left[ -\exp\left(-\gamma (X_T-q_0 \text{VWAP}_T) \right)\right] \le \sup_{v \in \mathcal{A}_{det}} \E \left[ -\exp\left(-\gamma (X_T-q_0 \text{VWAP}_T) \right)\right],$$ and the result follows since the converse inequality holds.
\begin{flushright}$\Box$\end{flushright}

Let us now come to the premium for guaranteed VWAP. The above results show that

$$ \sup_{v \in \mathcal{A}} \E \left[ -\exp \left(-\gamma \left(X_T-q_0 \text{VWAP}_T\right)  \right)\right] =-\exp\left(-\gamma \left( - \int_0^{q_0}F(z)dz - \mathcal{I}(q^*) \right)\right).$$

Hence, the premium for guaranteed VWAP is given by the following Theorem:

\begin{Theorem}[Premium for guaranteed VWAP]
$$\pi(q_0) = \int_0^{q_0}F(z)dz + \mathcal{I}(q^*)$$
\end{Theorem}

\section{Examples and Numerics}
\label{sect: examples and numerics}

We now turn to specific cases in which closed forms expressions can be obtained.

\subsection{VWAP strategies in the absence of permanent market impact}

An interesting case, within the deterministic framework, is the case where there is no permanent market impact (i.e. $f=0$). In that case indeed, we have that the optimal strategy is to follow the market volume curve. This is stated in the next Proposition:

\begin{Proposition}\label{noperm}
If $f = 0$ then:
$$q^*(t) = q_0 \left( 1 - \frac{Q_t}{Q_T}\right)$$
and
$$\pi(q_0) = Q_T L\left(\frac{q_0}{Q_T}\right)$$
\end{Proposition}

\textbf{Proof:}\\

Although the proof can be made directly using Jensen's inequality on $\mathcal{I}$, we prove the result using the hamiltonian system of Proposition \ref{prop:deterministic hamiltonian}. Considering the function $q$ defined by $q(t) = q_0 \left( 1 - \frac{Q_t}{Q_T}\right)$ and a constant function $p$ such that $p(t) \in \partial^- L\left(-\frac{q_0}{Q_T}\right)$, we have that $-\frac{q_0}{Q_T} = H'(p(t))$ so that:
$$\dot{q}(t) = -\frac{q_0 V_t}{Q_T} = V_t H'(p(t))$$
Also, straightforwardly:
$$\dot{p}(t) = 0 = \gamma \sigma^2 \left( q(t)-q_0\left( 1 - \frac{Q_t}{Q_T}\right)\right)$$
This proves that $q=q^*$, hence the first part of the result.\\

Coming to the premium, we have:

$$\pi(q_0) = \int_0^{q_0}F(z)dz + \mathcal{I}(q^*)= \int_0^T V_t L\left(-\frac{q_0}{Q_T}\right) dt =  Q_T L\left(\frac{q_0}{Q_T}\right).$$
\qed\\

The result of the above Proposition deserves a few comments. It states that, in the absence of permanent market impact, the optimal strategy is to have a trading curve that has the same shape as the relative market volume curve. One consequence is that, in practice, as far as the trading strategy is concerned, when permanent market impact can be (or is) ignored, one is interested in the estimation of the relative market volume curve and not in the estimation of the absolute value of the market volume. This remark is particularly important to understand why the deterministic market volume assumption provides a rather acceptable approximation of the real case for VWAP trading. Although the total volume traded over a day is highly variable, we know that the relative market volume curve is stable from one day to the other (except on witching days). Since the cumulated volume $Q_T$ only appears in the execution strategy through the ratio $\frac{Q_t}{Q_T}$, considering it stochastic does not play any role if we have already assumed that the relative market volume curve is deterministic.\footnote{In practice, most institutions use relative market volume curves computed in advance, based on historical data.} The value of $Q_T$ is however important to determine the premium of the guaranteed VWAP contract. Since this premium must be decided upon at time $t=0$, the value of $Q_T$ must be understood\footnote{We ignore here part of the risk.} as a forecast at time $0$ of the total market volume over the period $[0,T]$.\\

The above remarks only apply in the absence of permanent market impact but the assumption of a deterministic market volume is acceptable if we deviate slightly through the introduction of a small permanent market impact. Now, our goal is to understand the influence of permanent market impact in this framework, and the nature of the related deviation of the optimal strategy from the relative market volume curve.\\

\subsection{VWAP strategies when $V_t = V$, $f=k$ and $L(\rho) = \eta \rho^2$}

We now explore, for flat volume curves, the particular case where execution costs are quadratic, i.e. $ L(\rho)= \eta \rho^2$, and where permanent market impact is linear, i.e. $ f=k \in \R_+$, as in the initial Almgren-Chriss framework.

\begin{Proposition}\label{AC}
Assume $ V_t=V$, $f=k\ge 0$ and $L(\rho)=\eta \rho^2$. We have:
$$ q^*(t)=q_0\left(1-\frac{t}{T} \right)-q_0 w(t),$$
and
$$\pi(q_0) = \frac{\eta}{VT} q_0^2 + q_0^2 \int_0^T \left(\frac \eta V w'(t)^2 - \frac{k}{T} w(t) + \frac{\gamma \sigma^2}{2} w(t)^2\right) dt,$$ where $$w(t) = \frac{k}{\gamma \sigma^2 T} \sinh\left(\sqrt{\frac{\gamma \sigma^2 V }{2 \eta}} t\right)\left[ \tanh\left(\sqrt{\frac{\gamma \sigma^2 V }{2 \eta}} \frac{ T}{2} \right) -\tanh\left(\sqrt{\frac{\gamma \sigma^2 V }{2 \eta}}  \frac{t}{2}\right) \right].$$
\end{Proposition}

\textbf{Proof:}\\

We first see that $ H(p)= \frac{p^2}{4 \eta}$. Then, using Proposition \ref{prop:deterministic hamiltonian}, we obtain that:

$$\left\lbrace
\begin{array}{lcl}
\dot{p}(t) &=& \gamma \sigma^2 \left( q(t)-q_0\left( 1 - \frac{t}{T}\right)\right)+\frac{k q_0}{T}\\
\dot{q}(t) &=& \frac{V}{2\eta} p(t)
\end{array}\right. \ \ \ q(0)=q_0, \ \ q(T)=0.$$

This leads to:

$$ \ddot{q}(t)=\frac{\gamma \sigma^2 V }{2 \eta} \left( q(t)- q_0\left( 1 - \frac{t}{T}\right) \right)+\frac{kq_0 V}{2 \eta T}, \ \ q(0)=0, \ \ q(T)=0.$$

Therefore, the optimal strategy if of the form:

$$ q(t)= q_0\left( 1 - \frac{t}{T}\right) -\frac{k q_0}{\gamma \sigma^2 T } + \alpha \sinh\left(\sqrt{\frac{\gamma \sigma^2 V }{2 \eta}}t \right)+\beta \cosh\left(\sqrt{\frac{\gamma \sigma^2 V }{2 \eta}}t \right)$$

The initial and terminal conditions imply that:

$$ \alpha=\frac{k q_0}{\gamma \sigma^2 T} \frac{1}{\sinh\left(\sqrt{\frac{\gamma \sigma^2 V }{2 \eta}} T \right)} \left( 1 - \cosh\left(\sqrt{\frac{\gamma \sigma^2 V }{2 \eta}} T \right)\right), \ \ \beta=\frac{k q_0}{\gamma \sigma^2 T}.$$

We obtain then:

$$ q^*(t)=q_0\left( 1 - \frac{t}{T}\right)-\frac{k q_0}{\gamma \sigma^2 T} \left( 1 - \cosh\left(\sqrt{\frac{\gamma \sigma^2 V }{2 \eta}} t\right)\right)$$
   $$+\frac{k q_0}{\gamma \sigma^2 T}\frac{\sinh\left(\sqrt{\frac{\gamma \sigma^2 V }{2 \eta}} t \right)}{\sinh\left(\sqrt{\frac{\gamma \sigma^2 V }{2 \eta}} T \right)} \left( 1 - \cosh\left(\sqrt{\frac{\gamma \sigma^2 V }{2 \eta}} T \right)\right).$$

Since $ \frac{1-\cosh(x)}{\sinh(x)}=-\tanh(x/2)$, we obtain:

$$ q^*(t)=q_0\left(1-\frac{t}{T} \right)-\frac{k}{\gamma \sigma^2 T}q_0 \sinh\left(\sqrt{\frac{\gamma \sigma^2 V }{2 \eta}} t\right)\left[ \tanh\left(\sqrt{\frac{\gamma \sigma^2 V }{2 \eta}} \frac{ T}{2} \right) -\tanh\left(\sqrt{\frac{\gamma \sigma^2 V }{2 \eta}}  \frac{t}{2}\right) \right].$$

Coming to the premium, we have:

\begin{align*}
\pi(q_0)& = \int_0^{q_0}F(z)dz + \mathcal{I}(q^*)\\
& = \frac k2 q_0^2 + \int_0^T \left(\frac \eta V \dot{q}^*(t)^2 - \frac{k}{T} q_0 \left(q_0 - q^*(t)\right) + \frac{\gamma \sigma^2}{2} \left(q^*(t) - q_0\left(1 - \frac tT\right)\right)^2\right) dt\\
& = \frac k2 q_0^2 + \int_0^T \left(\frac \eta V \left(\frac {q_0} T + q_0 w'(t)\right)^2 - \frac{k}{T} q_0 \left(q_0 \frac t T + q_0 w(t)\right) + \frac{\gamma \sigma^2}{2} q_0^2 w(t)^2\right) dt\\
& = \frac \eta {VT} q_0^2 + \int_0^T \left(\frac \eta V q_0^2 w'(t)^2 + 2 \frac \eta {VT} q_0^2 w'(t)   - \frac{k}{T} q_0^2 w(t) + \frac{\gamma \sigma^2}{2} q_0^2 w(t)^2\right) dt\\
& = \frac{\eta}{VT} q_0^2 + q_0^2 \int_0^T \left(\frac \eta V w'(t)^2 - \frac{k}{T} w(t) + \frac{\gamma \sigma^2}{2} w(t)^2\right) dt.\\
\end{align*}
\begin{flushright}$\Box$\end{flushright}

This result permits to understand the role played by permanent market impact. We indeed have that $w \ge 0$ and therefore that the liquidation must occur more rapidly with the addition of permanent market impact. The rationale underlying this point is that the intermediary is going to pay $q_0 \text{VWAP}_T$ to the client and therefore he has an incentive to sell rapidly so that the price moves down, resulting in a lower VWAP. If $k$ is large, the optimal strategy may even be to oversell before buying back the shares so as to reduce the value of the VWAP (see below).\\

Coming to the premium for a guaranteed VWAP, it is straightforward to see that if $q(t) = q_0\left(1- \frac t T\right)$ the premium would be equal to $\frac{\eta}{VT} q_0^2$. Therefore, the reduction in the premium due to the use of the optimal strategy $q^*$ is given by

 $$-q_0^2 \int_0^T \left(\frac \eta V w'(t)^2 - \frac{k}{T} w(t) + \frac{\gamma \sigma^2}{2} w(t)^2\right) dt \ge 0.$$\\

In particular, in the limiting case $\gamma = 0$, corresponding to a risk neutral agent, we obtain the following straightforward formulas for the optimal strategy and the premium:

$$q^*(t) = q_0\left(1-\frac{t}{T} \right)\left(1-\frac{kV}{4\eta} t \right)$$
$$\pi(q_0) = \frac{\eta}{VT} q_0^2 - \frac{k^2 VT}{48\eta} q_0^2.$$

Several examples of optimal VWAP liquidation are given on Figure 1. We see that taking permanent market impact into account is important since the optimal trading curve may be really different from the simple trading curve obtained in Proposition \ref{noperm}.\\

\begin{figure}[h!]
\center
  \includegraphics[width=0.95\textwidth]{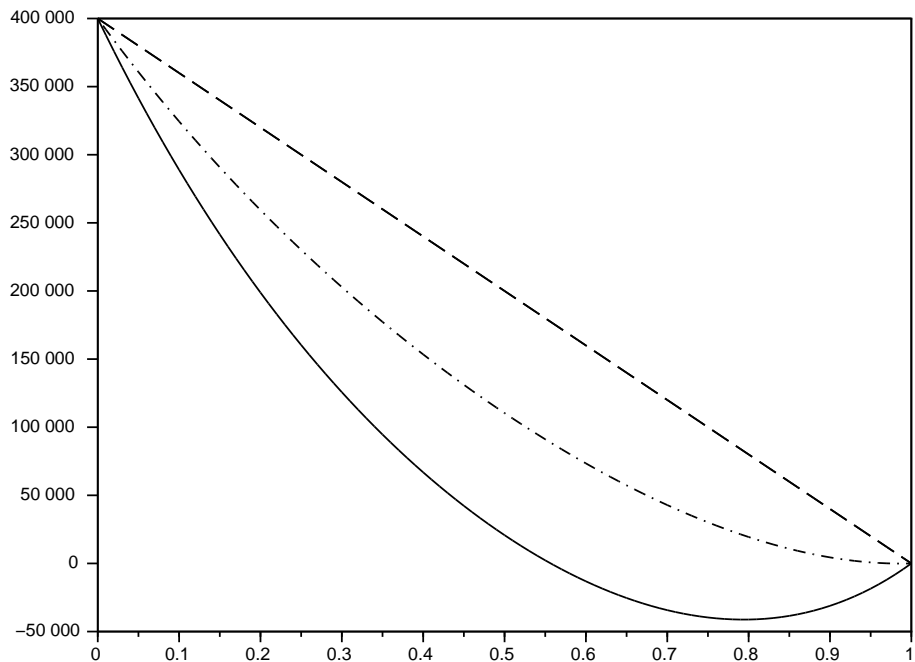}
  \caption{Examples of trading curves for a VWAP strategy. $S_0 = 50$, $q_0 = 400000$, $V=4000000$, $\sigma=0.45$, $\eta = 0.15$, $k = 5\times10^{-7}$, $T=1$ day. Plain line: $\gamma = 3\times 10^{-6}$. Dot-dashed line: $\gamma = 6\times 10^{-6}$. The dashed line corresponds to $q_0\left(1 - \frac tT\right)$.}
  \label{vwap}
\end{figure}

It is important to understand what is at play here. An agent willing to sell shares at a price close to the VWAP over a given period has usually two possibilities. He may call his favorite broker and ask for an agency VWAP order. In that case, the broker will try to sell shares as close as possible to the VWAP and the price obtained will be the price for the agent. In other words, the risk is borne by the agent. The other possibility is to enter a guaranteed VWAP contract. In that case, the price obtained by the agent will always be the VWAP. Our point is that the VWAP obtained by the agent in a guaranteed VWAP contract is not the same as the VWAP obtained on average through agency trades. In a guaranteed VWAP contract, the counterpart has indeed an incentive to sell more rapidly in order to push down the price and hence push down the VWAP. This is not market price manipulation, as the overall impact of the execution process would be the same independently of the trajectory. This is however a form of VWAP manipulation. Is the agent harmed? Somehow yes, although he gets its benchmark price. Nonetheless, since the counterpart of the contract makes money by selling more rapidly at the beginning of the execution process, he can redistribute part of it through a reduction of the premium...\\

To measure the difference between the naive strategy $q^{naive}(t) = q_0\left(1-\frac tT\right)$ and the optimal strategy, the best indicator is the premium of a guaranteed VWAP contract $\pi(q_0)$. We considered the same cases as on Figure 1, that is $S_0 = 50$, $q_0 = 400000$, $V=4000000$, $\sigma=0.45$, $\eta = 0.15$, $k = 5\times10^{-7}$, $T=1$ day, and two scenarios for the risk aversion parameter $\gamma$. The results on Figure 1 state that if the naive strategy was used, the minimum price of a guaranteed VWAP contract would be $3$ bps. However, when optimal strategies are used in the above examples, the intermediary would accept the contract without the payment of a premium, as the theoretical value of the premia are in fact negative.

\begin{table}[h!]
\begin{center}
\begin{tabular}{|c|c|c|}
  \hline
  & $\gamma = 3\times 10^{-6}$ & $\gamma = 6\times 10^{-6}$ \\
  \hline
  Premium with $q^{naive}$ & $3$ bps & $3$ bps \\
  \hline
  $ \frac{\pi(q_0)}{q_0 S_0}$ & $-3.2$ bps & $-1.3$ bps \\
  \hline
\end{tabular}
\end{center}
\caption{Premium of a guaranteed VWAP contract in the case of a naive strategy and in the case of the optimal strategy.}
\end{table}

\subsection{Numerical methods}

We treated above special cases for which closed form formulas could be obtained. In general, this is not the case and we present here a general method to approximate the solution of the Hamiltonian system. It is important to notice that the use of the Hamiltonian system is preferable to the use of Euler-Lagrange equation when it comes to numerics since the problem remains of order 1. The method we use to approximate the solution $(p,q)$ of the Hamiltonian system on the grid $\lbrace 0, \tau, \ldots, T=J \tau \rbrace$ is to apply a Newton method on the following nonlinear system of equations:

$$
\left\lbrace
\begin{array}{lcl}
p_{j+1} &=& p_j  + \tau \left(\gamma \sigma^2 \left( q_{j+1}-q_0\left( 1 - \frac{Q_{j+1}}{Q_J}\right)\right)+q_0\frac{V_{j+1}}{Q_J} f(|q_0 - q_{j+1}|)\right)\\
q_{j+1} &=& q_{j} + \tau V_{j+1} H'(p_j), \qquad 0 \le j < J
\end{array}\right.  $$
$$q_0=q_0, \ \ q_J=0.$$

To be more precise, we consider a first couple $(q^0,p^0) \in \R^{J+1} \times \R^{J+1}$ where:
 $$q^0_0 = q_0, \quad q^0_J = 0,$$
 $$p^0_0 = L'\left(\frac{1}{V_{j+1}} \frac{q^0_{j+1} - q^0_{j}}{\tau}\right),$$
 $$ p^0_{j+1} = p^0_j  + \tau \left(\gamma \sigma^2 \left( q^0_{j+1}-q_0\left( 1 - \frac{Q_{j+1}}{Q_J}\right)\right)+q_0\frac{V_{j+1}}{Q_J} f(|q_0 - q^0_{j+1}|)\right), \quad 0 \le j < J.$$

 Typically, we consider $q^0$ given by $q^0_j = q_0\left( 1 - \frac{Q_j}{Q_J}\right)$.\\

 Then, to go from $(q^n,p^n)$ to $(q^{n+1},p^{n+1})$ we consider the following method:

 $$q^{n+1} = q^{n} + q_0  \delta q^{n+1}, \qquad  p^{n+1} = p^{n} + \delta p^{n+1},$$

where $\left(\delta q^{n+1}, \delta p^{n+1}\right)$ solves the linear system:

$$
\left\lbrace
\begin{array}{lcl}
\delta p^{n+1}_{j+1} &=& \delta p^{n+1}_j  + \tau \left(\gamma \sigma^2 q_0   - \text{sign}(q_0 - q_{j+1}) (q_0)^2\frac{V_{j+1}}{Q_J} f'(|q_0 - q_{j+1}|)\right) \delta q^{n+1}_{j+1}\\
\delta q^{n+1}_{j+1} &=& \delta q^{n+1}_{j} + \tau \frac{1}{q_0} V_{j+1} H''(p^n_j) \delta p^{n+1}_j - \frac 1{q_0} \left( q_{j+1}^n - q^n_j - \tau V_{j+1} H'(p^n_j) \right)
\end{array}\right.  $$
$$\delta q^{n+1}_0=0, \ \ \delta q^{n+1}_J=0.$$

Two examples of the use of this method are shown on Figure 2 and Figure 3. The first one corresponds to $L(\rho) = \eta |\rho|^{1+\phi}$ with $\phi < 1$ and linear permanent market impact. The second one corresponds to $L(\rho) = \eta |\rho|^{1+\phi}$ with $\phi < 1$ and nonlinear permanent market impact of the form $f(q) = k\alpha q^{\alpha-1}$ with $\alpha \in (0,1)$.
\vspace{1cm}
\begin{figure}[htbp]
\center
  \includegraphics[width=0.75\textwidth]{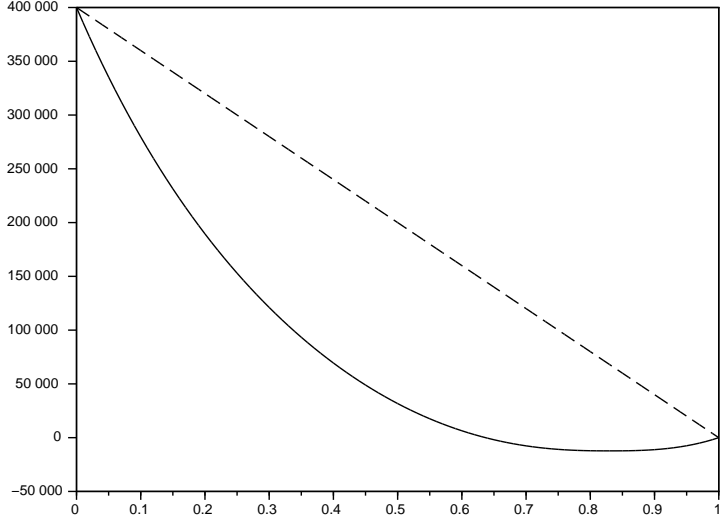}
  \caption{Examples of trading curves for a VWAP strategy (linear permanent market impact and nonlinear execution costs). $S_0 = 50$, $q_0 = 400000$, $V=4000000$, $\sigma=0.45$, $L(\rho) = \eta |\rho|^{1+\phi}$ with $\eta = 0.12$ and $\phi = 0.63$, $f=k = 5\times10^{-7}$, $T=1$ day. Plain line: $\gamma = 3\times 10^{-6}$. The dashed line corresponds to $q_0\left(1 - \frac tT\right)$.}
  \label{vwap2}
\end{figure}
\vspace{1cm}
\begin{figure}[htbp]
\center
  \includegraphics[width=0.75\textwidth]{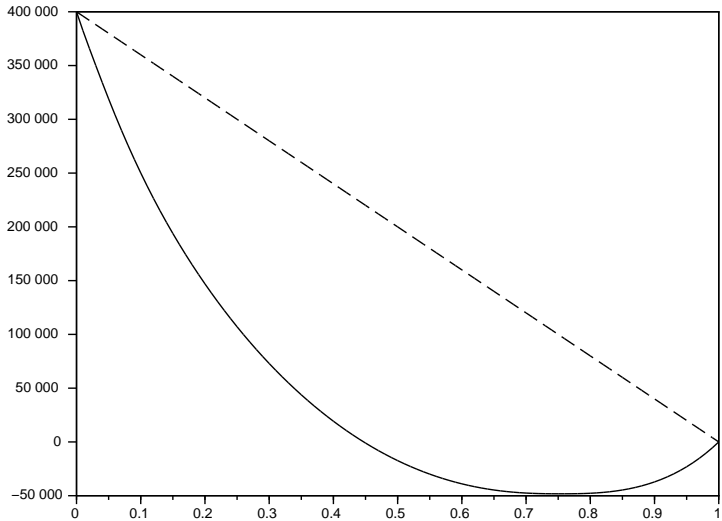}
  \caption{Examples of trading curves for a VWAP strategy (nonlinear permanent market impact and nonlinear execution costs). $S_0 = 50$, $q_0 = 400000$, $V=4000000$, $\sigma=0.45$, $L(\rho) = \eta |\rho|^{1+\phi}$ with $\eta = 0.12$ and $\phi = 0.63$, $f(q) = k\alpha q^{\alpha-1}$ with $k = 2.2\times10^{-4} $ and $\alpha = 0.6$, $T=1$ day. Plain line: $\gamma = 3\times 10^{-6}$. The dashed line corresponds to $q_0\left(1 - \frac tT\right)$.}
  \label{vwap3}
\end{figure}

\section{Going beyond the deterministic case}

In the previous sections of this article, we focused on the case of a deterministic volume curve. The reason for this is twofold. Firstly, it permits to understand the role played by permanent market impact in a tractable case. Secondly, it corresponds to the way VWAP strategies are often built in practice and we explained above why it was a good approximation. Practitioners usually compute relative market volume curves based on historical data and try to follow this curve to get a price as close as possible to the VWAP.\footnote{In practice, there are specific relative volume curves on special (witching) days. Some advanced desks also use several deterministic curves and may switch from one regime to another.}\\

We now briefly explore the case of stochastic market volume. The aim of this section is to provide a Hamilton-Jacobi-Bellman PDE to characterize the optimal liquidation strategy (that is no longer deterministic, and hence no longer a trading curve decided upon in advance, at time $0$) and the premium of a guaranteed VWAP contract. A similar approach, using stochastic optimal control, was adopted by Frei and Westray \cite{fw} in the mean-variance setup. However, in their paper, the initial filtration is augmented with the knowledge of the final volume and this makes their approach questionable for practical use.\\

In the model we consider, the instantaneous market volume is modeled by a simple stochastic process but it can be generalized to other processes. In fact, our main goal is to write the Hamilton-Jacobi-Bellman PDE characterizing the solution of our problem using as few variables as possible. If one considers indeed our problem with stochastic market volume in its initial form, 7 variables are necessary to describe the problem: the time $t$, the trader's inventory $q$, the asset market price $S$, the cash account $X$, the instantaneous market volume $V$, the cumulated market volume $Q$, and a variable linked to the VWAP, namely $\int_0^t V_s S_s ds$. Using two changes of variables, we manage to restrict the number of variables to 5. Classical numerical (PDE) methods may fail to approximate the solution of the PDE, but some probabilistic methods may be efficient (see \cite{glw} for robust methods in the case of problems in high dimension).\\

Coming to the model, we consider that the instantaneous volume is given by $ V_t=g(t) e^{\alpha B_t -\alpha^2 t/2}$, where $ g$ is $ C^1(\R,\R_+^*)$, where $ B$ is a Brownian motion independent of $ W$, and where $ \alpha>0$.  The dynamics of $ V$ is then given by the stochastic differential equation:
$$ \frac{dV_t}{V_t}= \frac{g'(t)}{g(t)} dt + \alpha dB_t.$$
$g(t)$ represents obviously the instantaneous market volume, on average, at time $t$. In Europe, it is a W-shaped curve with a peak corresponding to the opening of the US market.\\
\begin{Remark}
Other dynamics can be considered. The goal of this last section is not to consider the best possible model for volumes but rather to show how the complexity of the model can be reduced through changes of variables.\\
\end{Remark}
In order to consider a non-degenerated problem (one may alternatively use the stochastic target framework), we consider a slightly modified problem where $q_T$ is not forced to be equal to $0$. Rather than imposing $q_T=0$, we consider that, at time $T$, the remaining stocks are not liquidated at price $S_T$ but rather at price $S_T - K q_T$, where $ K$ is chosen positive and high enough to discourage the trader to keep a large position $q_T$ at time $ T$.\\

In this slightly modified framework, with computations similar to those of section \ref{subsect: some easy calculations}, we obtain the following:
\begin{align*}
X^v_T-q_0\text{VWAP}_T+q_TS_T-Kq^2_T= &-\int_0^{q_0} F(z) dz-\int_0^T V_t L\left( \frac{v_t}{V_t} \right) dt+q_0 \int_0^T \frac{V_t}{Q_T} F(q_0-q_t) dt\\
&+\sigma q_0 \int_0^T \left[ \frac{q_t}{q_0}-\left(1-\frac{Q_t}{Q_T} \right) \right]dW_t\\
&+\int_0^{q_T}F(z)dz-q_T F(q_0-q_T)-K q^2_T.
\end{align*}

Mathematically, it corresponds to a penalization at time $ T$ of the form $h(q) = -\int_0^qF(z) dz+qF(q_0-q)+Kq^2$, and we suppose that $K > 2 \limsup_{q \to +\infty}  \frac{F(q)}{q}$.\\

Now, we define the set of admissible strategies for all $ t\in[0,T]$ by:
$$ \mathcal{B}_t:= \left\{ v \in \mathcal{P}(t,T), \int_t^T |v_s|ds \in L^\infty  \right\}.$$

Our problem is then the following:
\begin{align}\label{maximisation volume sto}
\sup_{v \in \mathcal{B}_0} \E \left[ -\exp\left(-\gamma \mathcal{V}(v) \right)\right],
\end{align}
where:
$$\mathcal{V}(v) = -\int_0^T V_t L\left( \frac{v_t}{V_t} \right) dt+q_0 \int_0^T \frac{V_t}{Q_T} F(q_0-q_t) dt + \sigma q_0 \int_0^T \left( \frac{q_t}{q_0}-\left(1-\frac{Q_t}{Q_T} \right) \right)dW_t - h(q_T).$$

To solve this stochastic optimal control problem, we now introduce two processes $ \tilde{X}^v$ and $ \tilde{Y}^v$:
$$ d \tilde{X}_s^{t,x,v}=-V_s L \left( \frac{v_s}{V_s} \right) ds +(q_s-q_0) \sigma dW_s, \ s\in [t,T], \ X^{t,x,v}_t=x .$$
$$ d\tilde{Y}_s^{t,y,v}= V_s q_0 F(q_0-q_s) ds+Q_s \sigma q_0 dW_s, \ s\in[t,T], \ \tilde{Y}^{t,y,v}_t=y.$$
For any $ t \in [0,T]$, we define:
\begin{align}\label{maximisation volume sto dynamique}
U(t,x,y,q,Q,V):= \sup_{v\in \mathcal{B}_t} \E \left[ -\exp \left(-\gamma \left(\tilde{X}_T^{t,x,v}+\frac{\tilde{Y}_T^{t,y,v}}{Q_T}-h(q_T) \right)  \right)  \right].
\end{align}

This is the dynamic problem associated to \eqref{maximisation volume sto} and we observe that \eqref{maximisation volume sto} corresponds to $ U(0,0,0,q_0,0,V_0)$. We then have the following:

\begin{Theorem}
$ U $ is a viscosity solution of:
$$\left\lbrace
\begin{array}{lcl}
-\mathcal{L}U-\sup_{v \in \R } \big\{ -VL\left( \frac{v}{V} \right)\partial_{x} U-v \partial_q U \big\} = 0, \ \ \text{on} \ \in [0,T) \times \R^3\times \R_+ \times \R_+^{*}. \\
U(T,x,y,q,Q,V) = -\exp\left(-\gamma \left( x+\frac{y}{Q}-h(q) \right) \right).
\end{array}\right. $$
where the operator $ \mathcal{L}$ is defined by:
$$ \mathcal{L}:= \partial_t+(q-q_0)^2 \frac{\sigma^2}{2} \partial_{xx}+q_0V F(q_0-q) \partial_y+q_0^2Q^2 \frac{\sigma^2}{2} \partial_{yy}+V\partial_Q+V\frac{g'(t)}{g(t)} \partial_V +V^2 \frac{\alpha^2}{2}\partial_{VV}  . $$
\end{Theorem}

\textbf{Proof:}\\

It is straightforward to see that the value function is locally bounded. The result is then classically obtained by stochastic control technics. The required dynamic programming principle can indeed be deduced from the apparatus developed in \cite{prt}, and the viscosity subsolution and supersolution are obtained using \cite{bt}.
\begin{flushright}$\Box$\end{flushright}

It is noteworthy that we managed to remove the price $S$ from the state variables. We now remove $x$, using a change of variables. For that purpose, we introduce $\tilde{U}(t,y,q,Q,V) = e^{\gamma x} U(t,x,y,q,Q,V)$. Then we have:
\begin{Corollary}
 $\tilde{U}$ is a viscosity solution of:
$$\left\lbrace
\begin{array}{lcl}
-\tilde{\mathcal{L}}{\tilde{U}}-(q-q_0)^2 \frac{\sigma^2\gamma^2}{2} \tilde{U}-\sup_{v \in \R } \big\{ \gamma VL\left( \frac{v}{V} \right)\tilde{U}-v \partial_q \tilde{U} \big\} = 0, \ \ \text{on} \ \in [0,T] \times \R^3\times \R_+ \times \R_+^{*}. \\
\tilde{U}(T,y,q,Q,V) = -\exp\left(-\gamma \left( \frac{y}{Q}-h(q) \right) \right).
\end{array}\right. $$
with $ \tilde{\mathcal{L}}$ defined by:
$$ \tilde{\mathcal{L}}:= \partial_t +q_0V F(q_0-q) \partial_y+q_0^2Q^2 \frac{\sigma^2}{2} \partial_{yy}+V\partial_Q+V\frac{g'(t)}{g(t)} \partial_V +V^2 \frac{\alpha^2}{2}\partial_{VV}  . $$
\end{Corollary}

\textbf{Proof:}\\
The change of variables $\tilde{U}(t,y,q,Q,V) = e^{\gamma x} U(t,x,y,q,Q,V)$ being monotonically increasing, the result is obtained by straightforward computation.
\begin{flushright}$\Box$\end{flushright}

Now, we are going to prove a technical lemma in order to show that $U$ (or equivalently $\tilde{U}$) is always negative.
\begin{Lemma}
For any $ (t,x,y,q,Q,V)\in [0,T] \times \R^3\times \R_+ \times \R_+^{*}  $, we have:
$$ U(t,x,y,q,Q,V) <0.$$
\end{Lemma}
\textbf{Proof:}\\

Let us consider $(t,x,y,q,Q,V) \in [0,T] \times \R^3\times \R_+ \times \R_+^{*}$ and $ v\in \mathcal{B}_t$. We have by Jensen's inequality:

$$ \E \left[ -\exp\left(-\gamma \left(\tilde{X}_T^{t,x,v}+\frac{\tilde{Y}_T^{t,y,v}}{Q_T}-h(q_T) \right) \right)\right] \le  -\exp\left(-\gamma \mathcal{E}^v \right),$$

where

\begin{align*}
\mathcal{E}^v &= \E \left[\tilde{X}_T^{t,x,v}+\frac{\tilde{Y}_T^{t,y,v}}{Q_T}-h(q_T)\right]\\
&=\E \left[x + \frac{y}{Q_T} + \sigma q_0 \int_t^T \left( \frac{q_s}{q_0}-\frac{Q_T-Q_s}{Q_T} \right)dW_s\right.\\
& + \left.\int_t^T \left(-V_s L \left( \frac{v_s}{V_s} \right) + \frac{V_s}{Q_T} q_0 F(q_0-q_s) \right) ds  - h(q_T)\right].\\
\end{align*}

Since $v\in \mathcal{B}_t$, we have $\mathbb{E} \left[\int_t^T q_s dW_s \right]=0$. By independence of $V$ and $ W$, we also have that $\mathbb{E} \left[\int_t^T \frac{Q_T-Q_s}{Q_T} dW_s \right]=0$. Therefore:

$$\mathcal{E}^v = x + \E \left[\frac{y}{Q_T}\right] + \E \left[\int_t^T \left(-V_s L \left( \frac{v_s}{V_s} \right) + \frac{V_s}{Q_T} q_0 F(q_0-q_s) \right) ds - h(q_T)\right].$$

Since $f$ is nonincreasing on $\R_+$, there exists a constant $C \ge 0$ such that:
 $$ \forall q \in \R,\ \   F(q) \le C(1+q_+).$$
Now, since $ L$ is superlinear, there exists $B$ such that:
$$ \forall \rho \in \R,\ \  L(\rho) \geq -B + Cq_0 |\rho|.$$

This gives:

\begin{align*}
\mathcal{E}^v & \le x + \E \left[\frac{y}{Q_T}\right] + \E \left[\int_t^T \left(B - Cq_0 |v_s| + \frac{V_s}{Q_T} Cq_0 \left(1 + (q_0-q_s)_+\right) \right) ds - h(q_T)\right]\\
&\le x + \E \left[\frac{y}{Q_T}\right] +  BT + Cq_0  + \E \left[\int_t^T \left(- Cq_0 |v_s| + \frac{V_s}{Q_T} Cq_0 (q_0-q_s)_+ \right) ds - h(q_T)\right]\\
&\le x + \E \left[\frac{y}{Q_T}\right] + BT + Cq_0 + \E \left[Cq_0 \left((q_0-q_T)_+ - \frac{Q}{Q_T} (q_0-q)_+ \right) - h(q_T)\right]\\
& + \E \left[\int_t^T \left(- Cq_0 |v_s| - C q_0 \frac{Q_s}{Q_T} v_s \right) ds\right]\\
&\le x + BT + Cq_0 + \E \left[\frac{y}{Q_T}\right] +  \E \left[ Cq_0 (q_0-q_T)_+ - h(q_T) \right].\\
\end{align*}

Since $ Cq_0 (q_0-q)_+ - h(q) \underset{|q|\rightarrow +\infty} {\longrightarrow} -\infty$, we get that $\sup_{v \in \mathcal{B}_t}\E \left[ Cq_0 (q_0-q_T)_+ - h(q_T) \right] < +\infty$. The assumption on the market volume process $(V_t)_t$ then gives that $\E \left[\frac{1}{Q_T}\right]$ exists (and is independent of $v$). Putting these inequalities altogether, we get:

$$\sup_{v \in \mathcal{B}_t} \mathcal{E}^v < +\infty.$$

Therefore,

$$ U(t,x,y,q,Q,V) \le -\exp\left(-\gamma \sup_{v \in \mathcal{B}_t} \mathcal{E}^v \right) < 0.$$

\begin{flushright}$\Box$\end{flushright}

Now, since $\tilde{U}$ is never equal to $0$, we can consider the change of variables $ \tilde{U}(t,y,q,Q,V):=-\exp\left(\gamma \zeta(t,y,q,Q,V) \right)$. This new change of variables does not remove another variable but it has two related advantages. Firstly, $\zeta$ is in the same unit as the cash account. Hence, it takes values in a range that can be evaluated in advance. This is particularly important when it comes to numerics. Secondly, the premium $\pi(q_0)$ for a guaranteed VWAP contract is straightforwardly:
$$\pi(q_0)  = \int_0^{q_0} F(z) dz + \zeta(0,0,q_0,0,V_0).$$

Easy computations lead to the fact that $\zeta$ is a viscosity solution of:

$$\left\lbrace
\begin{array}{lcl}
\mathcal{L}_0{\zeta} + (q-q_0)^2 \frac{\sigma^2\gamma}{2} - V H(\partial_q \zeta) = 0, \ \ \text{on} \ [0,T)\times \R^2 \times \R_+^2. \\
\zeta(T,y,q,Q,V) = -\frac{Y}{Q}+h(q),
\end{array}\right. $$
where $H$ is the Legendre transform of $L$, and where the nonlinear operator $\mathcal{L}_0$ is defined by:
\begin{align*}
\mathcal{L}_0&= \partial_t +Vq_0 F(q_0-q) \partial_y  + \frac{q_0^2 \sigma^2 Q^2}{2} \left[ \gamma \left( \partial_y \right)^2 +\partial_{yy} \right]\\
 &+V \partial_Q + V \frac{g'(t)}{g(t)} \partial_V +\frac{\alpha^2 V^2}{2} \left[ \gamma \left( \partial_V \right)^2 + \partial_{VV}\right].
\end{align*}

\section*{Conclusion}

In this article we built a model to find the optimal strategy to liquidate a portfolio in the case of a guaranteed VWAP contract. When there is permanent market impact, we showed that the best strategy is not to replicate the VWAP but rather to sell more rapidly to push down the VWAP. Also, we use the indifference pricing approach to give a price to guaranteed VWAP contracts and we showed that taking into account permanent market impact permits, at least theoretically, to reduce substantially the price of guaranteed VWAP contracts. Finally, in the case of stochastic volumes, we developed a new model with only 5 variables and not 7 variables as in a naive approach.\\

\section*{Appendix A: $\text{VWAP}_T$ or $\text{VWAP}'_T$?}

In Section 2, we briefly discussed two alternative definitions of the VWAP over $[0,T]$:

$$\text{VWAP}_T:= \frac{\int_0^T S_t V_t dt}{Q_T},$$

and

$$\text{VWAP}'_T:= \frac{\int_0^T S_t (V_t + v_t) dt}{Q_T + q_0}.$$

In the former case, we exclude our own volume, while in the latter case, closer to the market definition, we include it. In fact, in both cases, our volume has an influence since the dynamics of the stock price depends on $v$. In this appendix, we are going to prove that using one or the other definition does not make any difference, up to a change in the function $f$ and in $\sigma$. This is in fact the consequence of the following Proposition:

\begin{Proposition}
For any $ v \in \mathcal{A}$,
\begin{align*}
X_T-q_0 \text{VWAP}'_T=&-\frac{Q_T}{Q_T+q_0}\int_0^{q_0}F(z)dz-\int_0^T V_t L\left(\frac{v_t}{V_t} \right)dt\\
&+q_0 \frac{Q_T}{Q_T+q_0} \int_0^T \frac{V_t}{Q_T} F(q_0-q_t) dt\\
&+ \sigma q_0 \frac{Q_T}{Q_T+q_0} \int_0^T \left(\frac{q_t}{q_0}-\left( 1-\frac{Q_t}{Q_T}\right) \right)dW_t.
\end{align*}
\end{Proposition}

\textbf{Proof:}\\

Let us integrate by parts in the definition of $\text{VWAP}'_T$:

\begin{align*}
\text{VWAP}'_T &= \frac{1}{Q_T+q_0} \int_0^T S_t (V_t + v_t) dt \\
&=   S_0 + \sigma \frac{1}{Q_T + q_0} \int_0^T\left( q_t + Q_T - Q_t\right) dW_t\\
&+ \frac{1}{Q_T + q_0} \int_0^T \left({Q_t}-{Q_T} \right)f(\left| q_0-q_t\right| )v_t dt - \frac{1}{Q_T + q_0} \int_0^T q_t f(\left| q_0-q_t\right| )v_t dt \\
&= S_0 + \sigma \frac{1}{Q_T + q_0} \int_0^T\left( q_t + Q_T - Q_t\right) dW_t\\
&- \frac{1}{Q_T + q_0} \int_0^T V_t F(q_0 - q_t) dt - \frac{1}{Q_T + q_0} \int_0^T q_t f(\left| q_0-q_t\right| )v_t dt\\,
&= S_0 + \sigma \frac{1}{Q_T + q_0} \int_0^T\left( q_t + Q_T - Q_t\right) dW_t\\
&- \frac{1}{Q_T + q_0} \int_0^T V_t F(q_0 - q_t) dt - \frac{1}{Q_T + q_0} \int_0^{q_0} F(z) dz.\\
\end{align*}

Since we still have that

$$ X_T=q_0 S_0-\int_0^{q_0}F(z)dz- \int_0^T V_t L\left( \frac{v_t}{V_t} \right)  dt + \int_0^T q_t \sigma dW_t,$$

we obtain:

\begin{align*}
X_T-q_0 \text{VWAP}'_T=&-\frac{Q_T}{Q_T+q_0}\int_0^{q_0}F(z)dz-\int_0^T V_t L\left(\frac{v_t}{V_t} \right)dt\\
&+q_0 \frac{Q_T}{Q_T+q_0} \int_0^T \frac{V_t}{Q_T} F(q_0-q_t) dt\\
&+ \sigma q_0 \frac{Q_T}{Q_T+q_0} \int_0^T \left(\frac{q_t}{q_0}-\left( 1-\frac{Q_t}{Q_T}\right) \right)dW_t.
\end{align*}

\begin{flushright}$\Box$\end{flushright}

The above Proposition states that $X_T-q_0 \text{VWAP}'_T$ is in fact equal to $X_T-q_0 \text{VWAP}_T$, had we replaced $f$ by $\frac{Q_T}{Q_T+q_0}f$ and $\sigma$ by $\frac{Q_T}{Q_T+q_0} \sigma$. Hence, if our volume represents a few percent of the market volume, using one definition of VWAP or the other does not really make a difference. Also, when the market volume process is assumed to be deterministic, the simple model we used over the course of this article leads to results that can be used when our volume is taken into account, if we apply the right multiplicative factors.\\
In the case of stochastic volume, instead of changing the values of the volatility and the market impact, we adapt the dynamic problem \eqref{maximisation volume sto dynamique} with the new adapted slippage, and we obtain the same PDE with a different terminal condition, which does not induce higher complexity.\\

\section*{Appendix B: Relative pricing of guaranteed VWAP contracts}

We now explore, for a deterministic volume curve, the case where the guaranteed VWAP contract is priced in basis point of the VWAP. In that case, the agent has to deliver $ q_0 (1-\lambda)VWAP_T$ to his client, for some $ \lambda$ decided upon at time $t=0$. With the notations of Section \ref{sect: general framework}, we are now facing the maximization problem:
\begin{align}\label{def maximisation prop}
\sup_{v\in \mathcal{A}} \E \left[ -\exp\left(-\gamma (X_T-q_0 (1-\lambda)\text{VWAP}_T) \right)\right].
\end{align}

To price the contract, we consider the value of $\lambda^*$ given by:\footnote{Jensen's inequality gives that we can bound ourselves to $\lambda\le 1$.}
$$ \lambda^*(q_0):=\sup \big\{ \lambda \in (-\infty,1) , \ \sup_{v\in \mathcal{A}} \E \left[ -\exp\left(-\gamma (X_T-q_0(1-\lambda) \text{VWAP}_T) \right)\right]  \le -1 \big\}.$$

\vspace{5mm}
\begin{Remark}
We can restrict as in Section \ref{sect: deterministic case} the strategies to the deterministic ones.
\end{Remark}

Using the same methodology as in Sections \ref{sect: general framework} and \ref{sect: deterministic case}, we obtain easily that, in this framework, the slippage is gaussian. The following Lemma is indeed the equivalent of Lemma \ref{Lemme 1 section 3}:

\vspace{5mm}
\begin{Lemma}
For any $ v\in \mathcal{A}_{det}$,  $ X_T-q_0(1-\lambda) \text{VWAP}_T$ is normally distributed with mean
$$ \lambda q_0 S_0-\int_0^{q_0}F(z)dz-\int_0^T V_t L\left(\frac{v_t}{V_t} \right)dt+q_0(1-\lambda)\int_0^T \frac{V_t}{Q_T} F(q_0-q_t) dt $$
and variance
$$  \sigma^2 q_0^2 \int_0^T \left(\frac{q_t}{q_0}-(1-\lambda)\left( 1-\frac{Q_t}{Q_T}\right) \right)^2dt .$$
\end{Lemma}

\vspace{5mm}
For $ \lambda \le 1$, our aim is then to compute:
\begin{align*}
h(\lambda):= \sup_{v \in \mathcal{A}_{det}}  & \Big\{ \lambda q_0 S_0-\int_0^{q_0}F(z) dz-\int_0^T V_t L\left(\frac{v_t}{V_t} \right)dt+q_0(1-\lambda)\int_0^T \frac{V_t}{Q_T} F(q_0-q_t) dt\\
&- \frac{\gamma}{2}\sigma^2 q_0^2 \int_0^T \left(\frac{q_t}{q_0}-(1-\lambda)\left( 1-\frac{Q_t}{Q_T}\right) \right)^2dt \Big\}.
\end{align*}
As in Section \ref{sect: deterministic case}, the maximization problem linked to $h(\lambda)$ clearly boils down to the minimization problem:
\begin{align}
\inf_{v \in \mathcal{A}_{det}} \displaystyle\int_0^T \left[  V_t L\left(\frac{v_t}{V_t} \right)-q_0(1-\lambda) \frac{V_t}{Q_T} F(q_0-q_t) + \frac{\gamma}{2}\sigma^2 q_0^2 \left(\frac{q_t}{q_0}-(1-\lambda) \left(1- \frac{Q_t}{Q_T}\right) \right)^2 \right] dt.
\end{align}

Then, defining $ \mathcal{I}^\lambda : AC_{q_0,0}(0,T) \to \mathbb{R} $ by:
\begin{align*}
\mathcal{I}^\lambda(q):=  \displaystyle\int_0^T \ell^\lambda(t,q(t),\dot{q}(t)) dt,
\end{align*}
where
$$ \ell^\lambda(t,q,v)):= V_t L\left(\frac{v}{V_t} \right)-q_0(1-\lambda)\frac{V_t}{Q_T} F(q_0-q)+\frac{\gamma}{2}\sigma^2 q_0^2\left(\frac{q}{q_0}-(1-\lambda)\left( 1-\frac{Q_t}{Q_T}\right) \right)^2,$$
we obtain:
\begin{Theorem}[Existence, uniqueness and Hamiltonian characterization of the maximiser]\label{existence lambda}
There exists a unique minimizer $ q^{\lambda,*}$ of $ \mathcal{I}^\lambda$ in $ AC_{q_0,0}(0,T)$. Moreover we have that $ q^{\lambda,*}(t) \le q_0$ for all $ t\in [0,T]$, and we have equivalence between the two following assertions:
\begin{itemize}
\item [$\rm{(i)}$] $ q=q^{\lambda,*}$;
\item [$\rm{(ii)}$] there exists $ p \in AC(0,T)$ such that for all $ t\in[0,T]$:
$$\left\lbrace
\begin{array}{lcl}
\dot{p}(t) &=& \gamma \sigma^2 \left( q(t)-q_0(1-\lambda)\left( 1 - \frac{Q_t}{Q_T}\right)\right)+q_0(1-\lambda)\frac{V_t}{Q_T}f(\left|q_0-q(t)\right| )\\
\dot{q}(t) &=& V_t H'(p(t)),
\end{array}\right.$$
with $ q(0)=0$ and $ q(T)=0$.
\end{itemize}
\end{Theorem}

The proof is similar to the proofs of Theorem \ref{existence} and Proposition \ref{prop:deterministic hamiltonian}.\\

We now end this section with the characterisation of $\lambda^*$:

\begin{Theorem}[Premium of the guaranteed VWAP contract]
\begin{align}\label{indif prop price}
 \lambda^*(q_0)=\sup \big\{\lambda \le 1, h(\lambda)\le 0 \big\},
\end{align}
where $ h$ verifies $ h(\lambda)=\lambda q_0 S_0-\int_0^{q_0}F(z) dz-\mathcal{I}^\lambda(q^{\lambda,*})$.
\end{Theorem}

This result is straightforward with the use of Theorem \ref{existence lambda}. To compute $\lambda^*$ numerically, we need to compute the values of the function $h$. This can be done through a numerical approximation of $q^{\lambda,*}$ using the same numerical methods as in Section 4.

\end{document}